\documentclass[journal=jacsat,manuscript=article]{achemso}

\usepackage[version=3]{mhchem} 
\usepackage{dirtytalk}
\usepackage{braket}
\usepackage{amsmath}
\usepackage{booktabs} 
\usepackage{textcomp}
\usepackage{xcolor}
\usepackage{tabto}
\usepackage{comment}
\usepackage{etoc}
\usepackage{url}
\usepackage{soul}
\usepackage{parskip}
\mciteErrorOnUnknownfalse

\SectionNumbersOn 
\PassOptionsToPackage{hyphens}{url}\usepackage{hyperref}
\makeatletter
\g@addto@macro{\UrlBreaks}{\UrlOrds}
\makeatother
\author{Shree Hari Sureshbabu}
\affiliation{School of Electrical and Computer Engineering, Purdue University, West Lafayette, IN, USA}
\author{Manas Sajjan}
\author{Sangchul Oh}
\affiliation{Department of Chemistry, Purdue University,
West Lafayette, IN, USA}
\affiliation{Department of Chemistry, Purdue University,
West Lafayette, IN, USA}
\author{Sabre Kais}
\affiliation{Department of Chemistry, Department of Physics and Astronomy, and Purdue Quantum Science and Engineering Institute, Purdue University, West Lafayette, IN, USA}
\email{kais@purdue.edu}

\title[An \textsf{achemso} demo]
  {Implementation of Quantum Machine Learning for Electronic Structure Calculations of Periodic Systems on Quantum Computing Devices}

\abbreviations{IR, NMR, UV}

\SectionNumbersOn
\begin{document}


\begin{abstract}
Quantum machine learning algorithms, the extensions of machine learning to quantum regimes, are believed to be more powerful as they leverage the power of quantum properties. Quantum machine learning methods have been employed to solve quantum many-body systems and have demonstrated accurate electronic structure calculations of lattice models, molecular systems, and recently periodic systems. 
A hybrid approach using restricted Boltzmann machines and a quantum algorithm to obtain the probability distribution that can be optimized classically is a promising method due to its efficiency and ease of implementation. Here we implement the benchmark test of the hybrid quantum machine learning on the IBM-Q quantum computer to calculate the electronic structure of typical 2-dimensional crystal structures: hexagonal-Boron Nitride and graphene. The band structures of these systems calculated using the hybrid quantum machine learning are in good agreement with those obtained by the conventional electronic structure calculation. This benchmark result implies that the hybrid quantum machine learning, empowered by quantum computers, could provide a new way of calculating the electronic structures of quantum many-body systems.
\end{abstract}

\section{Introduction}

Machine learning (ML) driven by big data and computing power has made a profound impact on various fields, including science and engineering~\cite{jordan2015machine}. Remarkably successful applications of machine learning range from image and speech recognition~\cite{he2016deep, sak2015learning} to autonomous driving~\cite{bojarski2016end}. The recent success of machine learning is mainly due to the rapid increase in classical computing power. This impact of ML has made it a useful tool to solve various problems in physical sciences~\cite{RevModPhys.91.045002}. Quantum computing is a new way of computation by harnessing the quantum properties such as the superposition and entanglement of quantum states. Some quantum algorithms run on quantum computers could solve the problems which are intractable by classical computers~\cite{arute2019quantum}. Recent progress in the development of Noisy Intermediate-Scale Quantum (NISQ) devices~\cite{preskill2018quantum}, makes it possible to run and test multiple quantum algorithms for various practical applications.

Quantum machine learning~\cite{Biamonte_2017}, the interplay of classical machine learning techniques with quantum computation, provides new algorithms that may offer tantalizing prospects to improve machine learning. At the same time, these techniques aid in solving the quantum many-body problems~\cite{lloyd2013quantum, rebentrost2014quantum, neven2008image, neven2008training, neven2009training, Das_Sarma_2019}. Using neural networks with a supervised learning scheme, Xu {\it et al.}~\cite{xu2018neural} have shown that measurement outcomes can be mapped to the quantum states for full quantum state tomography. Cong {\it et al.}~\cite{Cong_2019} have developed a quantum machine learning model motivated by convolutional neural networks, which makes use of ${\cal O}(\log(N))$ variational parameters for input sizes of $N$ qubits that allows for efficient training and implementation on near term quantum devices. 

It is important to solve many-body problems accurately for the advancement of material science and chemistry, as various material properties and chemical reactions are related to quantum many-body effects. Carleo and Troyer~\cite{carleo2017solving} introduced a novel idea of representing the many-body wavefunction in terms of artificial neural networks, specifically restricted Boltzmann machines (RBMs), to find the ground state of quantum many-body systems and to describe the time evolution of the quantum Ising and Heisenberg models. This representation was modified by Torlai {\it et al.}~\cite{torlai2018neural} for their purpose of quantum state tomography in order to account for the wavefunction's phase.

Quantum chemistry and electronic structure calculations using quantum computing are considered one of the first real applications of quantum computers ~\cite{aspuru2005simulated, doi:10.1002/9781118742631.ch01, peruzzo2014variational, kandala2017hardware, DASKIN201887}. Xia and Kais~\cite{Xia_2018} proposed a quantum machine learning method based on RBM to obtain the electronic structure of molecules. The traditional RBM was extended to three layers to take into account the signs of the coefficients for the basis functions of the wave function. This method was applied to molecular and spin-lattice systems. Recently, Kanno {\it et al.}~\cite{kanno2019manybody} have extended the method proposed by Xia and Kais by providing an additional unit to the third layer of an RBM in order to represent complex values of the wavefunctions of periodic systems. 

Since the discovery of graphene, it has sparked a huge interest due to its remarkable properties. Recently, there has been a lot of interest in studying graphene for quantum computing applications~\cite{joel2019coherent, calafell2019quantum}. Hexagonal Boron Nitride (h-BN) gained attention when it was shown that graphene electronics is improved when h-BN is used as a substrate for graphene~\cite{dean2010boron}. Of late the interest to study h-BN for quantum information has grown since it was discovered that the negatively charged Boron vacancy spin defects in h-BN display spin-dependent photon emission at room temperature~\cite{gottscholl2020initialization, gottscholl2020room, exarhos2019magnetic}. Hence, in addition to studying graphene, it is important to study h-BN as it is a potential candidate for creating spin qubits that can be optically initialized and readout.

In this paper, we implement the quantum machine learning method with a three-layered RBM along with a quantum circuit to sample the Gibbs distribution ~\cite{Xia_2018, kanno2019manybody} to calculate the electronic structure of periodic systems. Specifically, the implementation on NISQ devices is shown by modifying this quantum machine learning algorithm to run on an actual quantum computer. As the benchmark test, we demonstrate the performance of this algorithm first through the simulation of tight-binding and Hubbard Hamiltonians of hexagonal Boron Nitride and monolayer-graphene respectively, on the IBM quantum computing processors, which is done using the IBM quantum experience~\cite{aleksandrowicz2019qiskit}. The valance band of the 2-D honeycomb lattices is calculated using quantum machine learning methods on IBM-Q and the Qiskit simulator. As we shall see such valence band calculations on IBM-Q after employing a warm start and measurement error mitigation are shown to be in good agreement with the exact calculations.  

This paper is organized as follows. In Sec.~\ref{Sec:method}, the quantum machine learning method based on RBMs is introduced and implementation details are discussed. Sec.~\ref{Sec:results} presents the results of electronic structure calculations using quantum machine learning on the Qiskit simulator and IBM-Q quantum computers. Finally, the summary and discussion will be given in Sec.~\ref{Sec:conclusion}.

\section{Methodology}
\label{Sec:method}

In this section, we review the basic outline of the machine learning algorithm used and also discuss the implementation details
%
%
\subsection{Quantum Machine Learning Algorithm}

A quantum many-body state $\ket{\Psi}$ can be expanded in terms of the basis $\ket{\bf x}$, $\ket{\Psi} = \sum \Psi({\bf x})\ket{\bf x}$ where $\Psi({\bf x})$ is the wavefunction. Carleo and Troyer's {\cite{carleo2017solving}} method involved representing the trial wave function
$\Psi({\bf x};\theta)$ in terms of a neural network with parameters $\theta$ and to obtain the ground state by minimizing the expectation value of the Hamiltonian of a quantum many-body system, 
$E(\theta) = \bra{\Psi(\theta)}H\ket{\Psi(\theta)}$. This was shown to use lesser number of parameters compared to tensor-networks, indicating the efficiency of using such a representation.
More specifically, the ansatz of a trial wave function is given by
the marginal probability $P({\bf x} ;\theta)$ of a visible layer of the RBM,
$\Psi(\bf x;\theta) =\sqrt{P(\bf x;\theta)}$. While the learning of conventional RBMs is done by maximizing the likelihood function with respect to training data sets, the ground state of a neural network RBM state is obtained by minimizing the energy $E(\theta)$ using the stochastic optimization algorithm.

Xia and Kais~\cite{Xia_2018} introduced the third layer with a single unit to take into account the signs of the wavefunction and apply the quantum Restricted Boltzmann machine on actual quantum computers rather than the Monte-Carlo method on classical digital computers. This quantum machine learning algorithm was further extended to take into account the complex value of the wavefunction~\cite{kanno2019manybody}. However, implementation on an actual quantum computing processor was not shown, which would require multiple ancillary qubits as shown in this work.   

\begin{figure}[!htb]
    \centering
    \includegraphics[width=.55\textwidth]{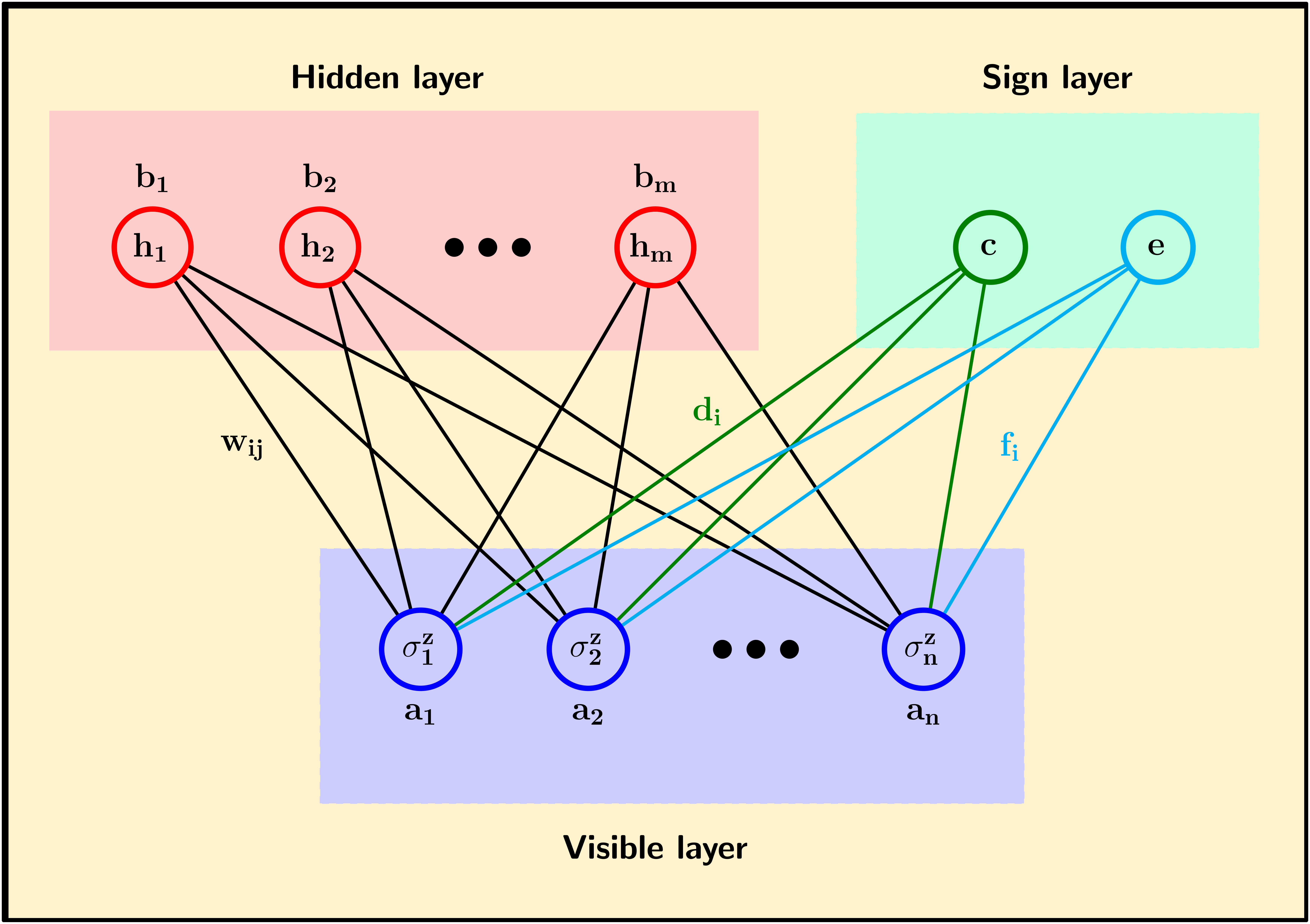}
    \caption{Restricted Boltzmann Machine used to calculate the electronic structure of periodic materials. Here, the sign layer consists of two units, one to account for the real part and the other for the complex part of the wavefunction.}
    \label{Fig:RBM}
\end{figure}

The RBM  we consider here consists of three layers: a visible layer, a hidden layer, and a complex layer, as shown in Fig.~\ref{Fig:RBM}. In contrast with the conventional RBMs with visible and hidden layers, the complex layer is added to take into account the real and imaginary values of the wavefunction of a quantum state. 

\noindent
The wavefunction of a periodic system can be expressed as:
\begin{equation}
\ket{\Psi} = \sum_{\bf x}\sqrt{P(\bf x)}s(\bf x)\ket{x} \,,
\end{equation}
where
\begin{align}
P({\bf x}) &= \frac{\sum_{\{h\}}e^{\sum_{i}a_i\sigma^z_i + \sum_{j}b_j h_j + \sum_{ij}w_{ij}\sigma^z_i h_j}}
{\sum_{\bf x'}\sum_{\{h\}}e^{\sum_{i}a_i\sigma^{z'}_i + \sum_{j}b_j h_j + \sum_{ij}w_{ij}\sigma^{z'}_i h_j}}\\[11pt]
s(\bf x) &= \tanh\left[(c + \sum_{i}d_i\sigma_i) + i(e + \sum_{i}f_i\sigma_i)\right]
\end{align}
Here $\sigma_i^z$ is the $z$-component of the Pauli operators at $i$,
$\ket{\bf x} = \ket{\sigma^z_1 \sigma^z_2 \sigma^z_3 ... \sigma^z_n}$ is the basis vector and the values that $\sigma^z_i$ and $h_j$ take are \{+1, -1\}. $a_i$, $b_i$, $c$, and $e$ denote the trainable bias parameters of the visible units, the hidden units, the unit representing the real part of the complex layer, and the unit representing the complex part of the complex layer, respectively. $w_{ij}$, $d_i$, and $f_i$ denote the trainable weights corresponding to the connections between $\sigma^z_i$ and $h_j$, $\sigma^z_i$ and the unit representing the real part of the complex layer, $\sigma^z_i$ and the unit representing the complex part of the complex layer, respectively. All the parameters are randomly initialized and the values of these random numbers range from -0.02 to 0.02.

\begin{figure}[!htb]
    \centering{
    \bigskip
    \includegraphics[width=0.90\textwidth]{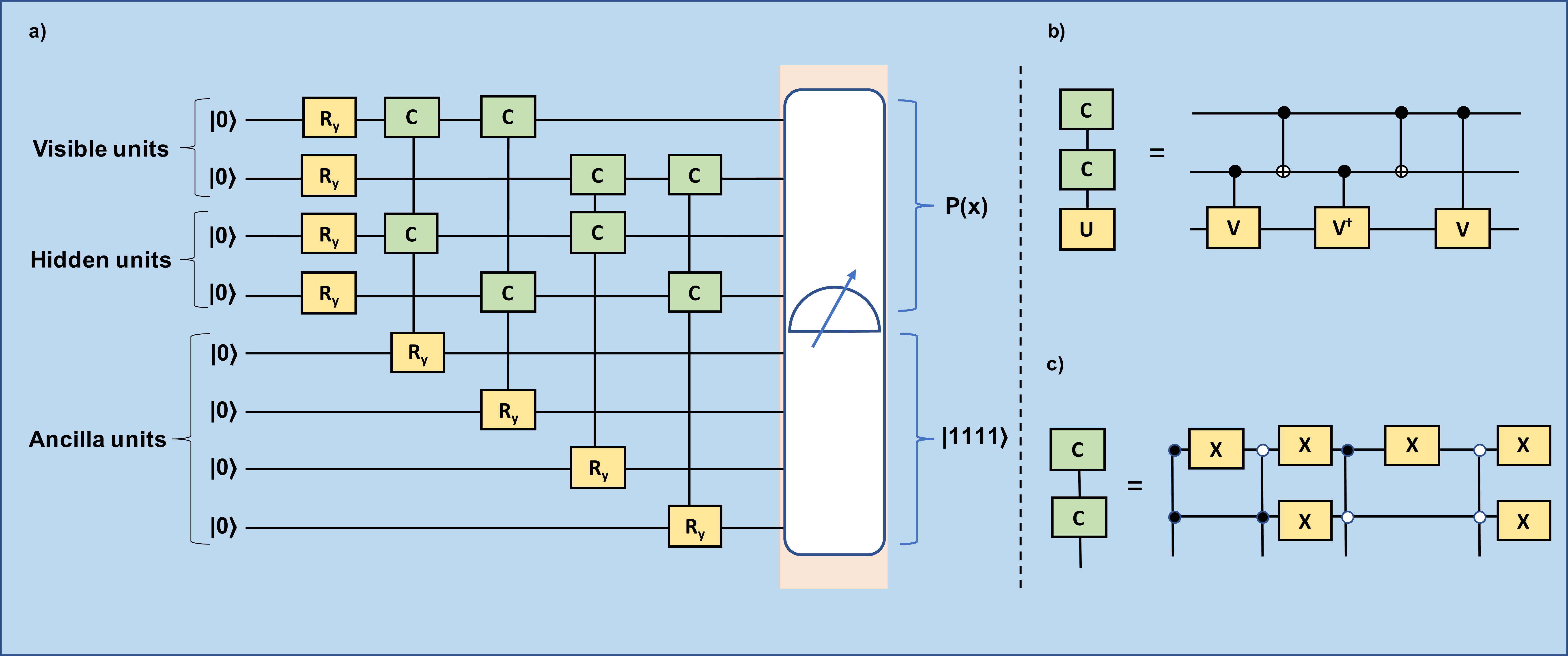}
    \bigskip}
    \caption{a) Quantum circuit to sample Gibbs distribution. This circuit consists of 2 visible units, 2 hidden units, and 4 ancilla qubits. $R_y$ represents the single qubit rotation, $C-C-R_y$ represents the controlled-controlled rotation, with visible and hidden units being the control qubits and ancilla qubit being the target qubit. After measurement, if the ancilla qubits are in $\ket{1111}$, only then the qubits correspnding to the visible and hidden units give the distribution ${P(\textbf{x})}$. b) Decomposition of the $C-C-R_y$ gate for $\ket{11}$. Here $U = V^2$ and this leads to choosing $V = R_y(\theta/2)$. c) $C-C-R_y$ conditioned by $\ket{00}$, $\ket{01}$, $\ket{10}$, and $\ket{11}$ can be achieved by implementing the circuit in this form.}
    \label{Q_circuit}
\end{figure}

\noindent

In order to obtain the probability distribution, the quantum circuit (shown in Fig.~\ref{Q_circuit}) is employed. The quantum circuit consists of a single qubit rotation ($R_y$) and a controlled-controlled rotation operations ($C-C-R_y$). The angle by which the $R_y$ operation rotates is determined by the visible and hidden bias parameters $a_i$ and $b_j$. The angle by which the $C-C-R_y$ operation rotates is determined by the weights connecting the visible and hidden layers $w_{ij}$.
For each combination of visible and hidden units, $y = \{\sigma^z, h\}$, in order to increase the probability of successful sampling, the distribution $Q(y)$ is sampled rather than $P(y)$~\cite{Xia_2018}. The two distribution functions $P(y)$ and $Q(y)$ are given by
\begin{align}
P(y) &= \frac{e^{\sum_{i}a_i\sigma^z_i + \sum_{j}b_j h_j + \sum_{ij}w_{ij}\sigma^z_i h_j}}{\sum_{y'}e^{\sum_{i}a_i\sigma^{z'}_i + \sum_{j}b_j h'_j + \sum_{ij}w_{ij}\sigma^{z'}_i h'_j}}\\[11pt]
Q(y) &= \frac{e^{\frac{1}{k}(\sum_{i}a_i\sigma^z_i + \sum_{j}b_j h_j + \sum_{ij}w_{ij}\sigma^z_i h_j)}}{\sum_{y'}e^{\frac{1}{k}(\sum_{i}a_i\sigma^{z'}_i + \sum_{j}b_j h'_j + \sum_{ij}w_{ij}\sigma^{z'}_i h'_j)}}
\end{align}
Here, $k$ is taken as $\max(\sum_{ij}\frac{|w_{ij}|}{2}, 1)$ \cite{Xia_2018}. This is done in order to make the lower bound of the probability of successful sampling a constant. If $k$ is taken to be 1, then the number of measurements required to get successful sampling becomes exponential. (See Supplementary Information).

The target qubits for the controlled-controlled Rotations are the ancilla qubits. Once all the rotations are completed, the ancilla qubits are measured. If the ancilla qubits are in $\ket{1}$, then the sampling is deemed successful. Then, the qubits corresponding to the visible and hidden units are measured to obtain the distribution $Q(y)$. Once the distribution $Q(y)$ is obtained, the probabilities are calculated to the power of $k$ and then normalized to get $P(y)$. With $P(y)$ computed through our QML algorithm and $s(y)$ computed classically, the wavefunction $\ket{\psi}$ is computed and through this the energy $E(\theta)$ is obtained. This value of $E(\theta)$ is optimized through gradient descent until the eigenvalue of the Hamiltonian is obtained. 

For this algorithm, the number of qubits required scales as $O(nm)$ and the complexity of the gates turns out to be $O(nm)$ for one sampling~\cite{Xia_2018}, where, $n$ is the number of visible units and $m$ being the number of hidden units.


\subsection{Implementation methods}

The developed quantum machine learning algorithm for calculating the band structures of h-BN and monolayer-graphene is executed using the following tools: 

(i) We start with the implementation of the algorithm classically. Classical simulation is performed to ensure the algorithm performs accurately. Here classical simulation implies that the gates were simulated on a classical computer.

(ii) Having ensured that the algorithm works when implemented classically, we move on to implementing it using Qiskit~\cite{aleksandrowicz2019qiskit}. Qiskit stands for IBM's Quantum Information Software Kit (Qiskit) and is designed to mimic calculations performed on a real noisy-intermediate scale quantum computing device using a classical computer. Specifically, we implemented the algorithm on the {\it qasm} backend, which is a high-performance quantum circuit simulator amenable to treat the errors (noise) associated with the implementation of the quantum circuit with appropriate customizable noise models. Essentially, the $qasm$ simulator is designed to replicate an actual noisy quantum device. Even if a custom noise is not chosen, depending on the circuit being executed the simulator automatically assumes a noise consistent with the hardware of the real device. The $C-C-R_y$ gate can be implemented by using {\textit{qiskit's}} multi-controlled y-rotation (mcry) operation, by specifying the control, target and ancillary qubits. The circuit is executed multiple times on the simulator each time culminating in the chosen set of measurements. The return values are the probabilities for observing the system in measurement basis states with statistical errors due to finite sampling.

(iii) We conclude our discussion by implementing and demonstrating the validity of our results using two actual IBM-Q quantum computers available. Qiskit's results in (i) are compared with those obtained from these real quantum devices. 

In the following section, we display the simulation results. The terms `RBM Value' and `Exact Value' stand for the values of valence band energies obtained from training our RBM and from exact diagonalization of the Hamiltonian, respectively.

Initializing the parameters of the RBM randomly can lead to the energies corresponding to certain $k$ points being stuck at local minima. To enhance the generalizing capability of a machine learning model, transfer learning technique has been successfully used. Recently, it has also been extended to the realm of quantum computing {\cite{mari2020transfer}}. However, in our case in order to improve the convergence, a method of warm starting is sufficient, wherein the parameters of a previously converged point are used to initialize the parameters of the current point of calculation. Noting that the band structure exists in a 4D space corresponding to energy as a function of $k_x, k_y$, and $k_z$, in this case too, if the optimization is performed such that the energy is minimized for every $(k_x, k_y, k_z)$ point, then the parameters of such a point in 4D space can be considered to improve the convergence of the other points.   

When implementing the algorithm on NISQ devices, we have to account for the noise that interferes with the accuracy of the results. In this work, we try to mitigate the errors that occur during measurement using \textit{Measurement Error Mitigation}. 

\begin{figure}[H]
    \centering
    \includegraphics[width=1.0\textwidth]{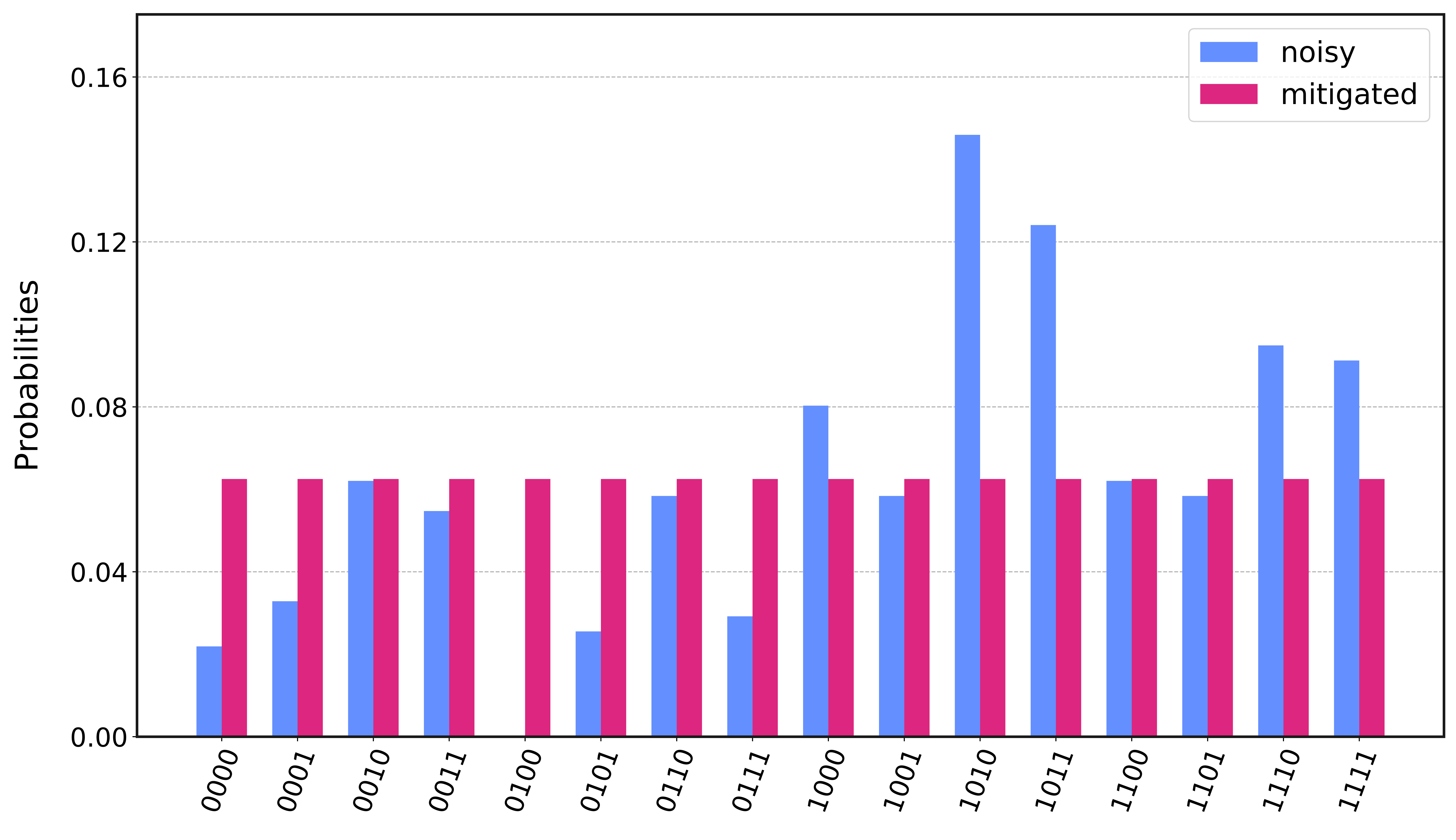}
    \caption{The probabilities of states with ancilla qubits being in $\ket{1111}$ for both the cases of with and without measurement error mitigation for the first iteration.}
\label{Measurement Error Mitigation}
\end{figure}

The counts corresponding to each state will not be definite as a result of noise. There will be a finite number of counts corresponding to the other basis states even when the measurement outcome is supposed to result in one. So the counts for each state can be written as a column vector and a matrix, called the calibration matrix, can be defined corresponding to the concatenation of all column vectors describing the counts for all the basis states. The least-squares method can now be used to get the error mitigated probabilities for each of the states by using the calibration matrix, the ideal state vector, and the noisy result that was obtained~\cite{aleksandrowicz2019qiskit}. An example of the probability distribution $Q(y)$ obtained with and without measurement error mitigation is shown in Fig.~\ref{Measurement Error Mitigation}.     

\section{Results and Discussion}
\label{Sec:results}

As a benchmark test of our quantum machine learning algorithm on existing IBM quantum computers, we calculate the electronic structures of two well-studied 2-dimensional periodic systems with hexagonal lattices namely Boron-Nitride and monolayer graphene. In this section, we discuss the results for each of the two systems.

\subsection{Band Structure of h-BN}
Hexagonal Boron nitride (h-BN) has a unit cell containing one B atom and another N atom. For h-BN, the levels involving the other valence orbitals, the $2s, 2p_x$, and $2p_y$, are either quite far above or far below the Fermi level. The conduction and valence bands, which are around the Fermi level, are formed from the $2p_z$ orbital and hence, a tight-binding Hamiltonian using the frontier $2p_z$ orbital and with third-nearest neighbor interaction on each of the two atoms of the unit cell is employed to obtain the electronic structures of the materials. Such a treatment affords the requisite dimensionality reduction as the number of qubits available on the IBM quantum computers is limited. Considering spin-degeneracy, the tight-binding Hamiltonian of the h-BN is thus given by a $4\times4$ Hermitian matrix (see Supplementary information). The number of visible units needed for the simulation is 2, and the number of hidden units is taken to be equal to the number of visible units. For quantum optimization, 2 qubits are used to represent the visible nodes and 2 qubits to represent the hidden nodes. In addition, 4 ancillary qubits are required (see Fig.~\ref{Q_circuit}). In total, the number of qubits required is equal to 8. The sampling of Gibb’s distribution is performed by applying the following sequences of quantum gates: 4 single-qubit rotation gates $(R_y)$, 16 controlled-controlled Rotation gates $(C-C-R_y)$, and 24 bit-flip $(X)$ gates, as illustrated in Fig.~\ref{Q_circuit}.

\begin{figure}[H]
    \centering
    \includegraphics[width=1.0\textwidth]{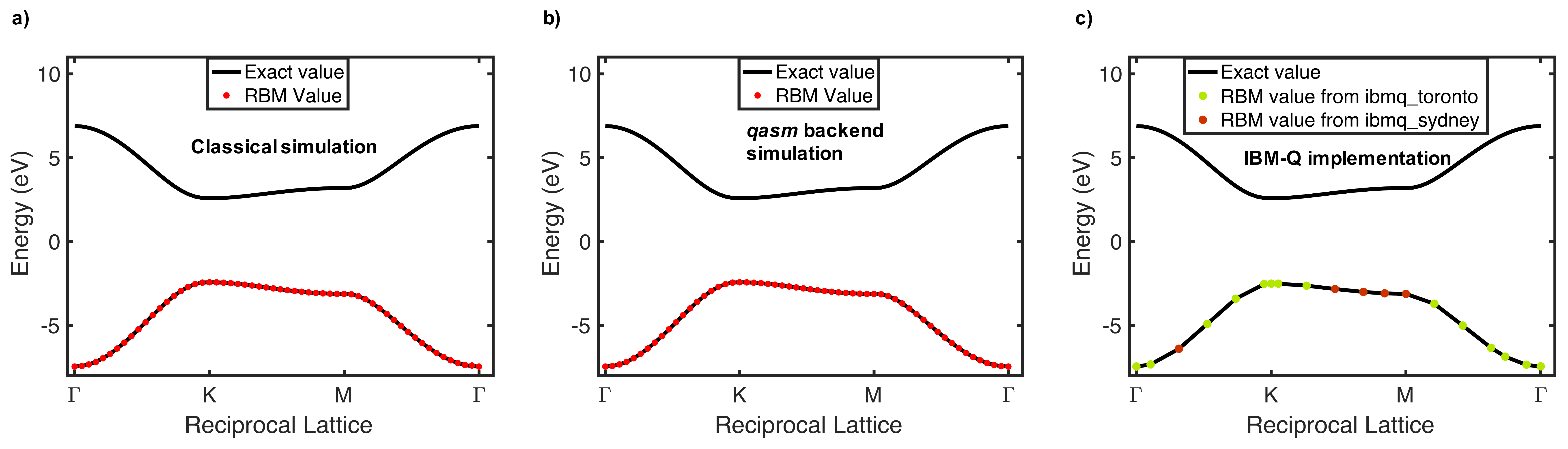}
    \caption{Band structures of h-BN calculated using (a) classical simulation with a warm start (red). The solid black curves show the valence and conduction bands from exact diagonalization. (b) 
    the $qasm$ backend simulation with the aid of a warm start (red).
    (c) The implementation the RBM sampling circuit on $ibmq\_toronto$ (green) and $ibmq\_sydney$ (red).}
\label{F:hBN_simulation}
\end{figure}

For h-BN band structure calculation, we start with the results of training RBM by implementing the gate-set (see Fig. \ref{Q_circuit}) classically and then on the Qiskit's quantum simulator, called the $qasm$ backend. 
Fig.~\ref{F:hBN_simulation} (a) shows the band structures of h-BN as a function of wave-vector amplitude sampled from the 1st Brillouin zone.
We overlay the valence band energies obtained from our RBM network on a classical computer with the exact diagonalization of the $4\times4$ tight-binding Hamiltonian (black curve). The two results are in excellent agreement. It must be noted that without a warm start, results may show deviations from the exact value at certain k-points as the optimization protocol may get locally trapped. However, the use of the warm starting technique eliminates such convergence issues. Fig.~\ref{F:hBN_simulation} (b) shows the band structure calculation of h-BN wherein for the RBM, the quantum gates are implemented on the Qiskit \textit{qasm} backend. For the sake of our simulations, no noise model was considered and the results obtained are just with statistical errors. Even in this case, if a warm start is provided, the quantum machine learning algorithm on the Qiskit \textit{qasm} simulator renders the exact valence band. In Fig.~\ref{F:hBN_simulation} (c) we show the implementation results for the valence band calculations using RBM wherein the gate-set is implemented on real IBM quantum devices, namely the \textit{ibmq\_toronto} and \textit{ibmq\_sydney}, both of which are 27 qubit devices. We see the results are in excellent agreement with the exact diagonalization when a warm start is provided along with \textit{Measurement Error Mitigation}.


\subsection{Band Structure of monolayer Graphene}

Much like h-BN, monolayer graphene also consists of two atoms in its unit cells. However, unlike the previous case, both the atomic centers are made up of carbon. Also, similar to h-BN, in the case of graphene, the levels involving the other valence orbitals, the $2s, 2p_x$, and $2p_y$, are either quite far above or far below the Fermi level. The orbital responsible for electrical conduction is just the $2p_z$ orbital and hence, a tight-binding Hamiltonian for the valence and conduction band with third-nearest neighbor interaction is constructed by taking into account the frontier $2p_z$ orbital on each of the carbons. The resultant matrix as before is a $4\times4$ matrix including spin-degeneracy(see Supplementary information). We introduce spin-spin interaction in graphene using the Fermi-Hubbard model with an onsite repulsion parameter $U$ between opposite spins.
In order to simulate graphene, the number of visible units and the number of hidden units is equal to 2. Therefore, 2 qubits to represent the visible nodes and 2 qubits to represent the hidden nodes, and in addition to that, 4 ancilla qubits are required. In total, the number of qubits required is equal to 8. The number of quantum gates required to sample Gibb’s distribution is 4 single qubit Rotation gates $(R_y)$, 16 Controlled-Controlled Rotation gates ($C-C-R_y$), and 24 Bit-flip (X) gates.

\begin{figure}[H]
    \centering
    \includegraphics[width=1\textwidth]{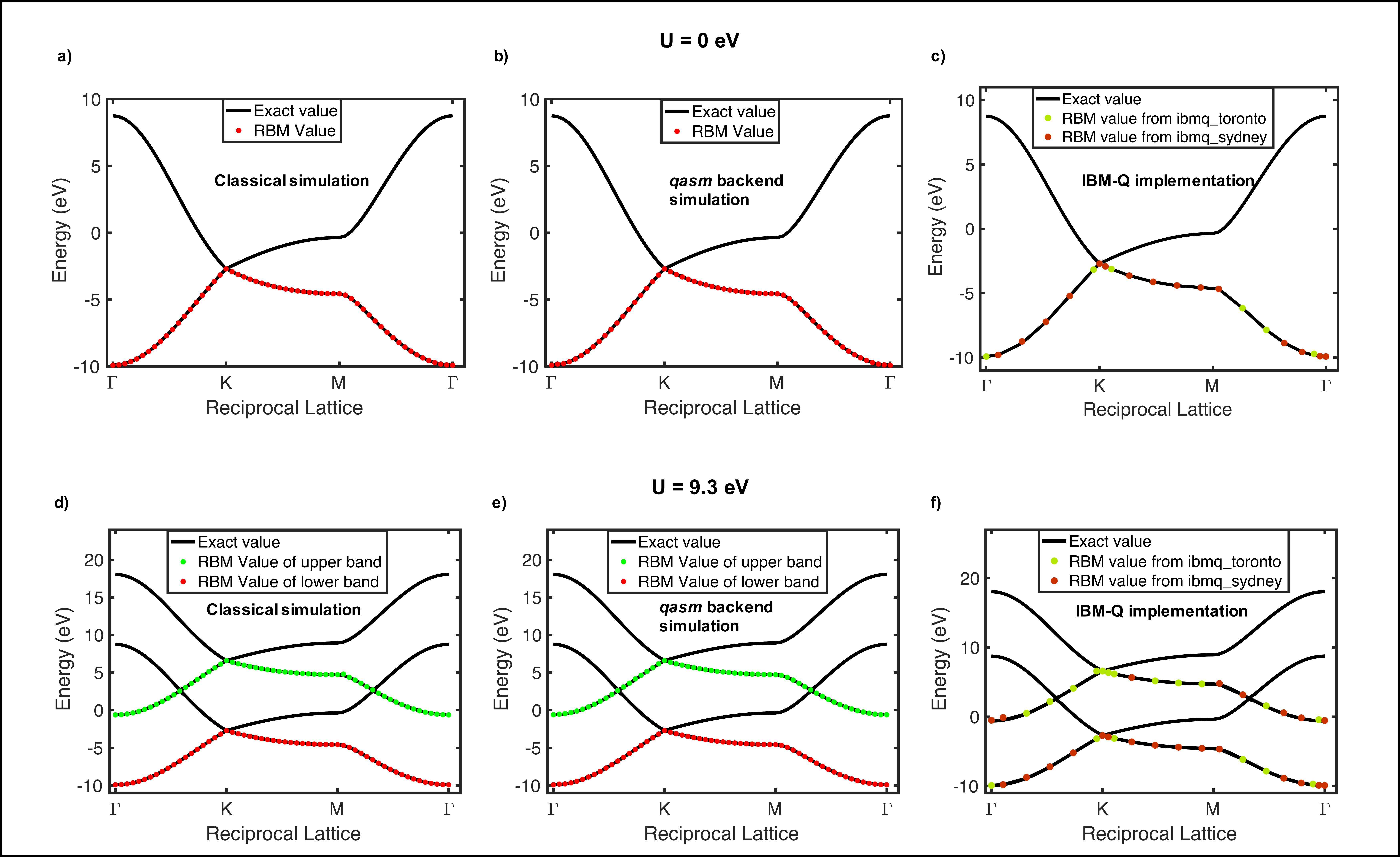}
    \caption{Band structure of the graphene for $U=0\,eV$ calculated using (a) classical simulation with a warm start (red). The solid black curves show the valence and conduction bands from exact diagonalization, (b) the $qasm$ backend simulation with the aid of a warm start (red). (c) Implementation on actual IBM computing devices. (d) Same as in (a) with Hubbard on-site interaction $U = 9.3\,eV$. The four bands correspond to the two non-degenerate spin-states for each of the valence and conduction bands in plot (a). (e) Same as in (b) with Hubbard on-site interaction $U = 9.3\,eV$. (f) Same as in (c) with Hubbard on-site interaction $U = 9.3\,eV$.}
\label{Graphene_U0}
\end{figure}

The band structures of monolayer graphene are calculated using the IBM Qiskit simulator and by running the QML algorithm on the IBM-Q quantum computers. Fig.~\ref{Graphene_U0} (a) shows the results for the band structures of graphene
at zero $U$ using the classical simulation. As before the results are overlayed on top of the eigenvalues obtained from exact diagonalization of the $4\times4$ Hamiltonian.
In Fig.~\ref{Graphene_U0} (b) we show the band structure of the graphene for $U=0$ calculated using the Qiskit qasm simulator. Finally, in Fig.~\ref{Graphene_U0} (c) we show the results of the quantum machine learning algorithm for calculation of the band structures of the graphene on IBM-Q quantum computers, the \textit{ibmq\_toronto}, and \textit{ibmq\_sydney}. Even for the case of graphene, the results are in good agreement with the exact diagonalization when a warm start is provided along with \textit{Measurement Error Mitigation}.

To show the band splitting for a non-zero on-site repulsion $U$, the Fermi level is shifted by a chemical potential $\mu = 15\;{\rm eV}$, which controls the filling of electrons. Fig.~\ref{Graphene_U0} (d-e)plots the band structures of graphene for $U = 9.3\; {\rm eV}$ obtained using the classical simulation, Qiskit $qasm$ backend, and the actual implementation on an IBM quantum computer. The RBM results are again in good agreement with that from exact diagonalization in all of the cases.



\subsection{Fidelity}

\begin{figure}[H]
    \centering
    \includegraphics[width=1.0\textwidth]{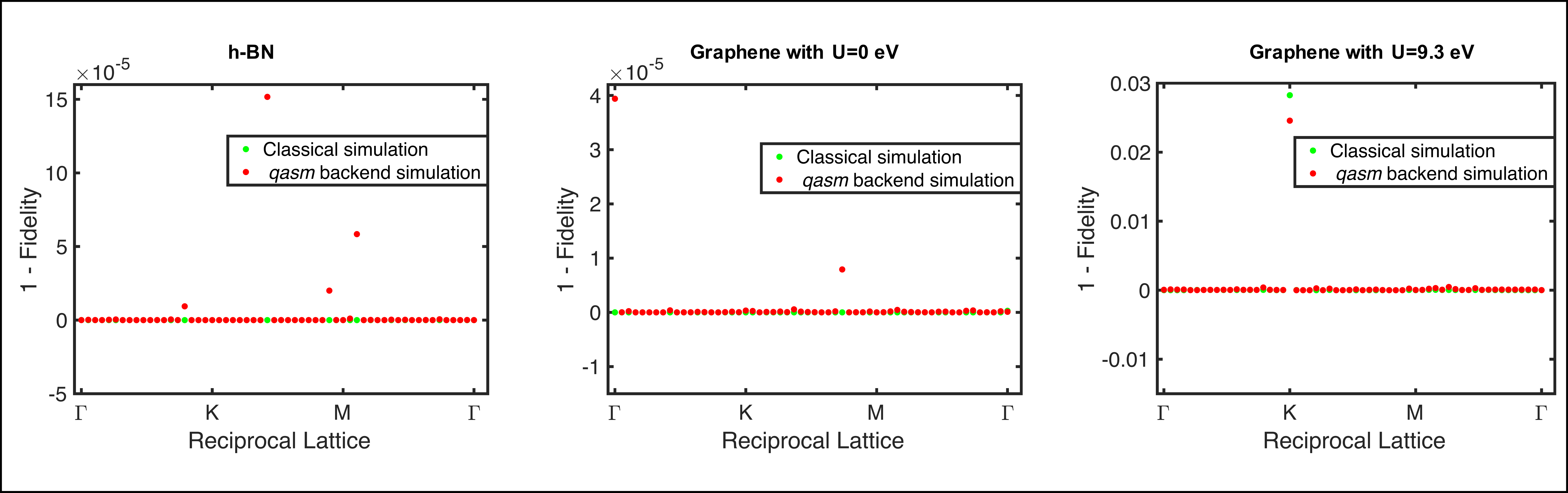}
    \caption{Error in fidelity $(1-F)$, are plotted as a function of the reciprocal lattice vector $(k)$ for classical simulation and $qasm$ backend.}
\label{Fidelity}
\end{figure}

To verify if the eigenstates provided by the QML algorithm match those obtained from exact diagonalization, the fidelity for each $k$ point is calculated. It can be seen from Fig.~\ref{Fidelity} that the error (1-Fidelity) is very small for classical simulation and simulation on the $qasm$ backend for both the materials. The fidelity is calculated as follows:
\begin{equation}
Fidelity = |\braket{\Psi|\Phi}|^2  \nonumber \\ 
\label{fidelity_eq}
\end{equation}
where,
$\ket{\Psi}$ is the eigenvector obtained from QML and $\ket{\Phi}$ is the eigenvector obtained from exact diagonalization.

\section{Conclusion}
\label{Sec:conclusion}

The primary goal of this study was to examine the performance of an RBM on a NISQ device in order to calculate the electronic structure of materials. In this work, the materials that were taken under consideration were hexagonal Boron Nitride (h-BN) and monolayer Graphene, both of which are two-dimensional solids. A tight-binding and a Hubbard Hamiltonian were constructed for h-BN and graphene respectively. By using an RBM and a quantum circuit to sample Gibbs distribution, the valence band energies for each of the two materials were obtained. In the case of graphene, the simulations were performed first for the case when the Hubbard $U$ is equal to 0 and then for the case of non-zero $U$. The band splitting for the case of non-zero $U$ was also shown. The simulations for both, graphene and h-BN were done using IBM's $qiskit$ framework as well as on real IBM quantum computing platforms.

Implementing RBM classically can either use Maximum-likelihood based gradient descent (which has a time complexity that is exponential in the size of the smallest layer){\cite{fischer2014training}} or Contrastive Divergence using Gibbs sampling, a Markov Chain Monte Carlo (MCMC) method (which is a more efficient approach) to estimate the gradients {\cite{carreira2005contrastive}}. The time complexity for training an RBM in the classical case scales as $O(N)$ , where $N$ is the size of the training data, while the implementation of RBM on a quantum computer has been shown to have quadratic speed-ups {\cite{wiebe2016quantum}}. Also, computing the ground state of a given Hamiltonian using exact diagonalization has a complexity of $\approx q^{3}$, where q is the dimension of the column space of a given matrix {\cite{harris2020array}}. However, setting $k= max(1,\frac{|w_{ij}|}{2})$ provides a constant lower bound in the probability of successful sampling and thus the complexity for one iteration scales as $O(mnN)$, where N is the number of successful sampling required to get the distribution P(x). 

The current quantum machine learning method could calculate only on the ground state energy of the periodic systems, i.e., the valence band, an extension is needed to treat systems with multiple valence bands {\cite{cerasoli2020quantum}} or to procure higher order energy bands. This can be done by sampling the orthogonal subspace of the previously computed valence band. Also, to calculate the transition matrix elements, the valence and conduction Bloch wavevectors should be obtained. The expectation value of an operator with respect to the ground state may be calculated using the Hellmann-Feynman method~\cite{Oh2009}. Here, the effect of noise on quantum machine learning is not fully explored, while the Qiskit qasm simulator and IBM-Q noisy quantum computers show the effect of noise on quantum optimization.
With the field of quantum computing developing rapidly, the curiosity of combining machine learning and quantum computing has led to very interesting researches. With the development of quantum computers and their capability to scale very fast, quantum machine learning can prove to be useful in not only electronic structure methods, but also as a significant tool in developing new materials and understanding complex phenomena.


\section{Data and model availability}
The input Hamiltonians corresponding to h-BN and graphene can be found in section 2 of the Supplementary Information. Data will be made available upon reasonable request to the corresponding author. The codes associated with the classical simulation, simulation on the $qasm$ backend, and the implementation on IBM's quantum computing devices will be made available with the corresponding author upon reasonable request.

\begin{acknowledgement}
We would like to thank Dr. Ruth Pachter, AFRL, for many useful discussions. AFRL support is acknowledged. We acknowledge the National Science Foundation under award number 1955907. This material is also based upon work supported by the U.S. Department of Energy, Office of Science, National Quantum Information Science Research Centers. We also acknowledge the use of IBM-Q and thank them for the support. The views expressed are those of the authors and do not reflect the official policy or position of IBM or the IBM Q team.

\end{acknowledgement}



\bibliography{references}


\newpage
\begin{center}
\textbf{\Large Supplementary Information for Implementation of Quantum Machine Learning for Electronic Structure Calculations of Periodic Systems on Quantum Computing Devices}
\end{center}

\setcounter{equation}{0}
\setcounter{figure}{0}
\setcounter{table}{0}
\setcounter{page}{1}
\setcounter{section}{0}

\section{General Tight-Binding Hamiltonian for a system of two sublattices}
We begin our discussion with a general tight-binding (TB) Hamiltonian for a system of consisting of unit cells of two conjoint sublattices (say A and B) with one atom per sublattice. For simplification we shall consider the case where each atom contributes only one orbital even though this restriction can be relaxed in a straight-forward extension. Our TB Hamiltonian in the basis of the contributing orbitals is:

\begin{eqnarray}
\rm{H} &=& \sum_{m,n,\sigma, \sigma^{\prime}} \epsilon^A_{m,n,\sigma, \sigma^{\prime}} a_{m\sigma}^\dagger a_{n\sigma^{\prime}} + \sum_{m,n,\sigma, \sigma^{\prime}} \epsilon^B_{m,n,\sigma, \sigma^{\prime}} b_{m\sigma}^\dagger b_{n\sigma^{\prime}}  \nonumber \\ 
&+& \sum_{m,n,\sigma, \sigma^\prime} t_{m,n,\sigma,  \sigma^\prime} (a_{m,\sigma}^\dagger b_{n, \sigma^\prime} + h.c.)  \label{Ham_eq}
\end{eqnarray}
where $\epsilon^A_{m,n,\sigma, \sigma^{\prime}}$, $\epsilon^B_{m,n,\sigma, \sigma^{\prime}}$ are the interaction matrix elements within each of the respective sub-lattices (either A or B) and $t_{m,n,\sigma,  \sigma^\prime}$ (assumed to be real) denotes the hopping interaction between the two-sublattices. $a_{m\sigma}^\dagger$ creates an electron in the $mth$ atom (also $mth$ orbital) with spin $\sigma$ in sublattice A. Similar definition holds also for $b_{m\sigma}^\dagger$ except it caters to the B sublattice. The following properties of these operators will be very useful later

\begin{eqnarray}
\{a^\dagger_{m\sigma}, a_{n\sigma^\prime}\} = \delta_{m,n,\sigma, \sigma^\prime} \label{prop1}\\
\{b^\dagger_{m\sigma}, b_{n\sigma^\prime}\} = \delta_{m,n,\sigma, \sigma^\prime} \label{prop2}\\
\{a^\dagger_{m\sigma}, a^\dagger_{n\sigma^\prime}\} = \{b^\dagger_{m\sigma}, b^\dagger_{n\sigma^\prime}\} = 0 \label{prop3}\\
\{b^\dagger_{m\sigma}, a^\dagger_{n\sigma^\prime}\} = \{b^\dagger_{m\sigma}, a_{n\sigma^\prime}\} = 0 \label{prop4}\\
a_{n\sigma}|0\rangle = 0 \label{prop5}\\
b_{n\sigma}|0\rangle = 0 \label{prop6}
\end{eqnarray}

Using Eq.\ref{prop4} is equivalent to assuming that the overlap metric between the sublattices A and B is identity. Now since Eq.\ref{Ham_eq} is banded, to afford dimensionality reduction and ease of diagonalization let us define Fourier transform of the operators 
$a^\dagger_{n\sigma^\prime}$ and 
$b^\dagger_{n\sigma^\prime}$ as follows:

\begin{eqnarray}\label{FT_operators}
c^\dagger_{k\sigma} = \frac{1}{\sqrt{N}}\sum_m e^{i\bf{k}\cdot\bf{R_{mA}}} a^\dagger_{m\sigma} \label{FT_operators_1} \\
c^\dagger_{k\sigma^\prime} = \frac{1}{\sqrt{N}}\sum_m e^{i\bf{k}\cdot\bf{R_{mA}}} a^\dagger_{m\sigma^\prime} \label{FT_operators_2}\\
d^\dagger_{k\sigma} = \frac{1}{\sqrt{N}}\sum_m e^{i\bf{k}\cdot\bf{R_{mB}}} b^\dagger_{m\sigma} \label{FT_operators_3}\\
d^\dagger_{k\sigma^\prime} = \frac{1}{\sqrt{N}}\sum_m e^{i\bf{k}\cdot\bf{R_{mB}}} b^\dagger_{m\sigma^\prime} \label{FT_operators_4}
\end{eqnarray}

\noindent
where $\bf{R}_{mB}$ and $\bf{R_{mA}}$ are real-space lattice vectors of the two sublattices and $\bf{k}$ is the wavevector that belongs to the 1st Brillouin zone of the corresponding reciprocal lattice. Using Eq.\ref{FT_operators_1}, \ref{FT_operators_2}, \ref{FT_operators_3}, \ref{FT_operators_4} and properties listed in Eq.\ref{prop1}, \ref{prop2}, \ref{prop3}, \ref{prop4}, \ref{prop5}, \ref{prop6}, it is now possible to construct matrix elements of the following forms:

\begin{itemize}
    \item $\langle 0 |c_{k\sigma^\prime} \rm{H} c^\dagger_{k\sigma}|0\rangle$
    \begin{eqnarray}
    \langle 0 |c_{k\sigma^\prime} \rm{H} c^\dagger_{k\sigma}|0\rangle&=& \frac{1}{N}\sum_{m,n,\sigma_1, \sigma_2}\sum_{p.q} e^{i\bf{k}\cdot(\bf{R_{qA}}-\bf{R_{pA}})} \langle a_{p\sigma^\prime} a^\dagger_{m\sigma_1}a_{n\sigma_2}a^\dagger_{q\sigma}\rangle \epsilon^A_{m,n,\sigma_1, \sigma_2} \nonumber \\
    &=& \frac{1}{N}\sum_{m,n,\sigma_1, \sigma_2}\sum_{p.q} e^{i\bf{k}\cdot(\bf{R_{qA}}-\bf{R_{pA}})} \delta_{nq}\delta_{\sigma\sigma_2}\delta_{mp}\delta_{\sigma^\prime\sigma_1} \epsilon^A_{m,n,\sigma_1, \sigma_2}\nonumber \\
    &=& \frac{1}{N}\sum_{p,q} e^{i\bf{k}\cdot(\bf{R_{qA}}-\bf{R_{pA}})}\epsilon^A_{p,q,\sigma, \sigma^{\prime}} \label{elem1}
    \end{eqnarray}
    
\item $\langle 0 |d_{k\sigma^\prime} \rm{H} d^\dagger_{k\sigma}|0\rangle$ 
\begin{eqnarray}
\langle 0 |d_{k\sigma^\prime} \rm{H} d^\dagger_{k\sigma}|0\rangle &=& \frac{1}{N}\sum_{p,q} e^{i\bf{k}\cdot(\bf{R_{qB}}-\bf{R_{pB}})}\epsilon^B_{p,q,\sigma, \sigma^{\prime}}  \label{elem2}
\end{eqnarray}

\item $\langle 0 |c_{k\sigma^\prime} \rm{H} d^\dagger_{k\sigma}|0\rangle$ 
\begin{eqnarray}
\langle 0 |c_{k\sigma^\prime} \rm{H} d^\dagger_{k\sigma}|0\rangle&=& 
\frac{1}{N}\sum_{m,n,\sigma_1, \sigma_2}\sum_{p.q} e^{i\bf{k}\cdot(\bf{R_{qB}}-\bf{R_{pA}})} \langle a_{p\sigma^\prime} a^\dagger_{m\sigma_1}b_{n\sigma_2}b^\dagger_{q\sigma}\rangle t_{m,n,\sigma_1, \sigma_2} \nonumber \\
    &=& \frac{1}{N}\sum_{m,n,\sigma_1, \sigma_2}\sum_{p.q} e^{i\bf{k}\cdot(\bf{R_{qB}}-\bf{R_{pA}})} \delta_{nq}\delta_{\sigma\sigma_2}\delta_{mp}\delta_{\sigma^\prime\sigma_1} t_{m,n,\sigma_1, \sigma_2}\nonumber \\
    &=& \frac{1}{N}\sum_{p,q} e^{i\Vec{\bf{k}}\cdot(\bf{R_{qB}}-\bf{R_{pA}})}t_{p,q,\sigma, \sigma^{\prime}} \label{elem3}
\end{eqnarray}
\end{itemize}

With these matrix elements, we can cast the Hamiltonian in the basis of operators defined in Eq.\ref{FT_operators_1}, \ref{FT_operators_2}, \ref{FT_operators_3}, \ref{FT_operators_4}. To proceed further we need to now specialize to the exact geometry of the lattice which shall be discussed in the next section.

\section{Honeycomb lattices: Graphene and h-BN}

\begin{figure}[H]
    \centering
    \includegraphics[width=0.9\textwidth]{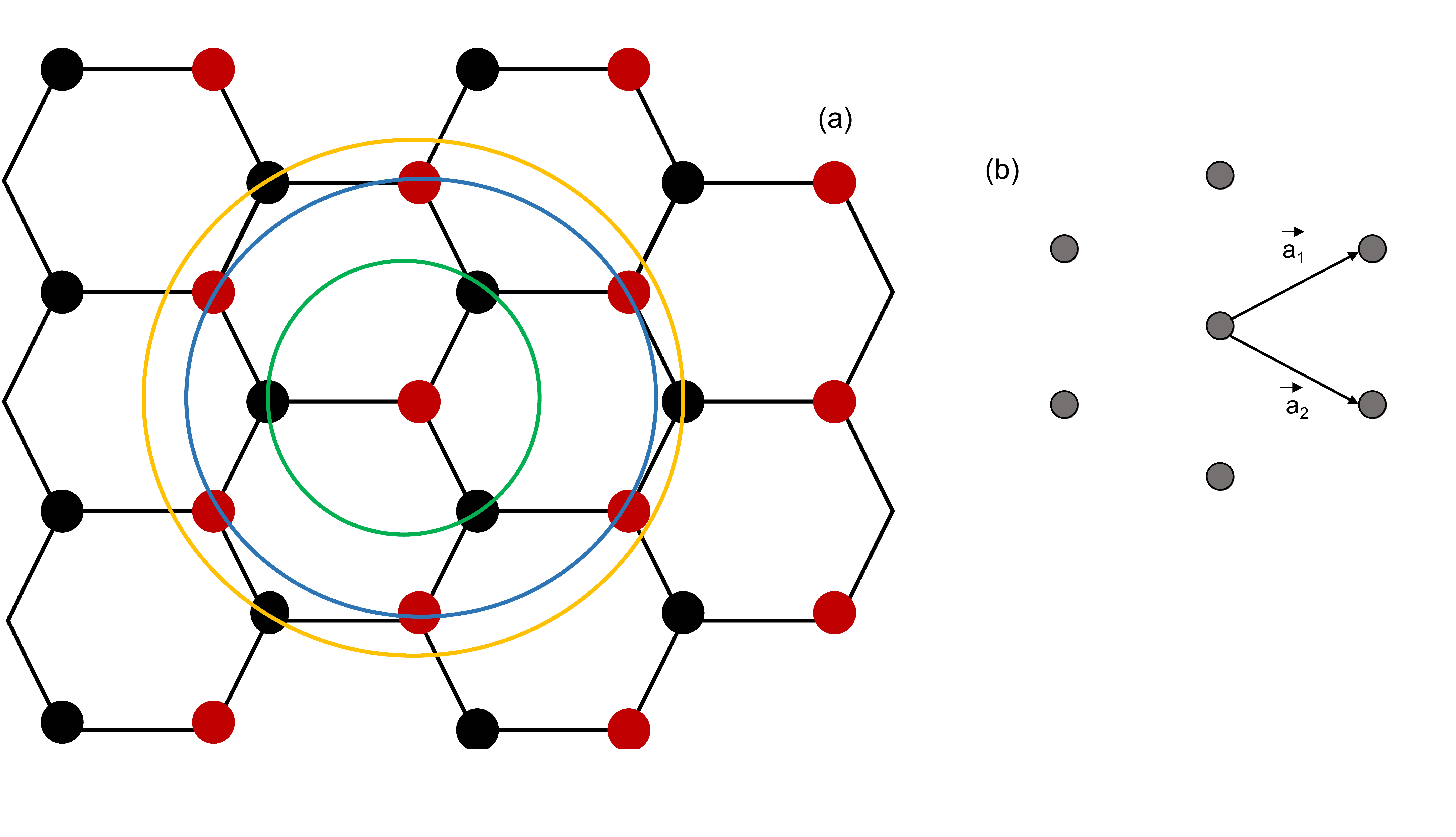}
    \caption{ (a) Structure of the honeycomb lattice. The green circle shows the nearest neighbors (three in this case), the blue circle shows the second-nearest neighbors (six in this case), and the orange circle shows the third-nearest neighbors (three in this case). (b) The unit vectors $\bf{a_1}$, $\bf{a_2}$ of the real space lattice are indicated. }
    \label{hex_lattice}
\end{figure}

Using the matrix elements derived in Eq.\ref{elem1}, \ref {elem2}, \ref{elem3} we can now deduce the Hamiltonian used in this work for graphene and h-BN upto third nearest neighbor interaction. Both graphene and h-BN possesses the similar lattice structure, a representative prototype of which is given in Fig.\ref{hex_lattice}. The real-space lattice unit vectors are $\bf{a_1}$, $\bf{a_2}$ are also displayed. The primitive vectors of the real lattice of are given by
${\bf a}_1 = a\left(\frac{\sqrt{3}}{2}, \frac{1}{2}\right)\,, \quad
{\bf a}_2 = a\left(\frac{\sqrt{3}}{2}, -\frac{1}{2}\right)$
where $a =|{\bf a}_1| = |{\bf a}_2| = 2.47$ {\AA} is the lattice constant for h-BN and $a = 2.55$ {\AA} for monolayer graphene.

\subsection{Nearest-neighbor interaction}
For nearest-neighbor interaction only in hexagonal honeycomb lattices, it is easy to appreciate from the geometry in Fig.\ref{hex_lattice}(a) that atoms in A sublattice share a vertex with those at sublattice B only and vice versa. So the following substitutions need to be made 
\begin{itemize}
    \item $\epsilon^A_{p,q,\sigma, \sigma^{\prime}} = \epsilon^A_p \delta_{pq}\delta_{\sigma \sigma^\prime}$ 
    
    Substituting in Eq.\ref{elem1} we get
    \begin{eqnarray}
    \langle 0 |c_{k\sigma^\prime} \rm{H} c^\dagger_{k\sigma}|0\rangle = \sum_p \frac{\epsilon^A_p}{N} \delta_{\sigma, \sigma^\prime}
    \end{eqnarray}
    
    \item $\epsilon^B_{p,q,\sigma, \sigma^{\prime}} = \epsilon^B_p \delta_{pq}\delta_{\sigma \sigma^\prime}$ 
    
    Substituting in Eq.\ref{elem2} we get
    \begin{eqnarray}
    \langle 0 |d_{k\sigma^\prime} \rm{H} d^\dagger_{k\sigma}|0\rangle = \sum_p \frac{\epsilon^B_p}{N} \delta_{\sigma, \sigma^\prime}
    \end{eqnarray}
    
    \item $t_{p,q,\sigma, \sigma^{\prime}} = t_1  \:\:\rm{if}\:\: \bf{R_{qB}} = \bf{R_{pA}} + \bf{d_1}, \bf{R_{pA}} + \bf{d_2}, \bf{R_{pA}} + \bf{d_3}$\\
    Substituting in Eq.\ref{elem3} we get
    \begin{eqnarray}
    \langle 0 |c_{k\sigma^\prime} \rm{H} d^\dagger_{k\sigma}|0\rangle &=&
    \delta_{\sigma \sigma^\prime} e^{i\bf{k}\cdot(-\bf{d_1})} t_1 (1+ e^{i\bf{k}\cdot(\bf{d_2}-\bf{d_1})} + e^{i\bf{k}\cdot(\bf{d_3}-\bf{d_1})}) \nonumber \\
    &=& \delta_{\sigma \sigma^\prime} e^{i\bf{k}\cdot(-\bf{d_1})} t_1 (1+ e^{i\bf{k}\cdot\bf{a_1}} + e^{i\bf{k}\cdot\bf{a_2}}) \nonumber 
    \end{eqnarray}
    
\end{itemize}
This form of the matrix elements have actually been deduced before in \cite{kundu2009tight, dresselhaus1998physical}. From geometry the nearest neighbor length vectors are 
$\bf{d_1} = (0, \frac{1}{\sqrt{3}})a,\bf{a_2} = \bf{d_2}-\bf{d_1}, \bf{a_3} = \bf{d_3}-\bf{d_1}$. 


\subsection{Second-nearest neighbor interaction}

From Fig.\ref{hex_lattice}(a) it is clear that every next nearest neighbor of atom A is also only atom A and vice-versa of atoms of B sublattice as well. Inclusion of second nearest neighbor interaction thus only modifies the matrix elements $\langle 0 |c_{k\sigma^\prime} \rm{H} c^\dagger_{k\sigma}|0\rangle$ and 
$\langle 0 |d_{k\sigma^\prime} \rm{H} d^\dagger_{k\sigma}|0\rangle$ only. Let us look at each of them

\begin{itemize}
    \item $\epsilon^A_{p,q,\sigma, \sigma^{\prime}} = \epsilon^A_p \delta_{pq}\delta_{\sigma \sigma^\prime}$ \\
    = $t_2$  if $\bf{R_{qA}}=\bf{R_{pA}} \pm \bf{a_1}$,$\bf{R_{qA}}=\bf{R_{pA}} \pm \bf{a_2}$, $\bf{R_{qA}}=\bf{R_{pA}} \pm \bf{(a_1 - a_2)}$. \\
    Substituting in Eq.\ref{elem1} we get
    \begin{eqnarray}
    \langle 0 |c_{k\sigma^\prime} \rm{H} c^\dagger_{k\sigma}|0\rangle &=& \sum_p (\frac{\epsilon^A_p}{N})\delta_{\sigma, \sigma^\prime} + 
    t_2 (e^{-i\bf{k}\cdot\bf{a_1}}+ e^{-i\bf{k}\cdot\bf{a_2}} + e^{i\bf{k}\cdot\bf{a_1}} +
    e^{i\bf{k}\cdot\bf{a_2}}  \nonumber \\
    &+& e^{i\bf{k}\cdot\bf{(a_1-a_2)}}+ e^{-i\bf{k}\cdot\bf{(a_1-a_2)}})\delta_{\sigma, \sigma^\prime}
    \end{eqnarray}
    
\item  $\epsilon^B_{p,q,\sigma, \sigma^{\prime}} = \epsilon^B_p \delta_{pq}\delta_{\sigma \sigma^\prime}$ \\
    = $\Tilde{t_2}$  if $\bf{R_{qB}}=\bf{R_{pB}} \pm \bf{a_1}$,$\bf{R_{qB}}=\bf{R_{pB}} \pm \bf{a_2}$, $\bf{R_{qB}}=\bf{R_{pB}} \pm \bf{(a_1 - a_2)}$. \\  
    Substituting in Eq.\ref{elem2} we get
    \begin{eqnarray}
    \langle 0 |d_{k\sigma^\prime} \rm{H} d^\dagger_{k\sigma}|0\rangle &=& \sum_p (\frac{\epsilon^B_p}{N})\delta_{\sigma, \sigma^\prime} + 
    \tilde{t_2} (e^{-i\bf{k}\cdot\bf{a_1}}+ e^{-i\bf{k}\cdot\bf{a_2}} +
    e^{i\bf{k}\cdot\bf{a_1}} +
    e^{i\bf{k}\cdot\bf{a_2}}  \nonumber \\
    &+& e^{i\bf{k}\cdot\bf{(a_1-a_2)}}+ e^{-i\bf{k}\cdot\bf{(a_1-a_2)}})\delta_{\sigma, \sigma^\prime} 
    \end{eqnarray}
    
    \item $\langle 0 |c_{k\sigma^\prime} \rm{H} d^\dagger_{k\sigma}|0\rangle = 
    \delta_{\sigma \sigma^\prime} e^{i\bf{k}\cdot(-\bf{d_1})} t_1 (1+ e^{i\Vec{\bf{k}}\cdot\Vec{\bf{a_1}}} + e^{i\bf{k}\cdot\bf{a_2}})$
\end{itemize}
The matrix elements of the type $\langle 0 |c_{k\sigma^\prime} \rm{H} d^\dagger_{k\sigma}|0\rangle$ do not change at all and is equal to the value obtained in the nearest-neighbor case.

\subsection{Third-nearest neighbor interaction}
It is evident from Fig.\ref{hex_lattice}(a) that third-nearest neighbor interaction only interconnects of the atoms in A and B sublattices only and hence matrix elements of the kind
$\langle 0 |c_{k\sigma^\prime} \rm{H} d^\dagger_{k\sigma}|0\rangle$ will be exclusively changed while elements of the kind $\langle 0 |d_{k\sigma^\prime} \rm{H} d^\dagger_{k\sigma}|0\rangle$ and $\langle 0 |c_{k\sigma^\prime} \rm{H} c^\dagger_{k\sigma}|0\rangle$ will involve participation upto second nearest neighbor only.

\begin{itemize}
    \item $t_{p,q,\sigma, \sigma^{\prime}} = t_1  \:\:\rm{if}\:\: \bf{R_{qB}} = \bf {R_{pA}} + \bf{d_1}, \bf {R_{pA}} + \bf{d_2}, \bf {R_{pA}} + \bf {d_3}$ and \\
    = $t_3 \:\:\:\rm{if}\:\: \bf {R_{qB}} = \bf {R_{pA}} \pm (\bf {a_1-a_2}), \bf{R_{pA}} + \bf{a_2+a_1}$ \\
 Substituting these in Eq.\ref{elem3} we get 
 \begin{eqnarray}
 \langle 0 |c_{k\sigma^\prime} \rm{H} d^\dagger_{k\sigma}|0\rangle &=& 
 t_1(1 + e^{i\bf{k}\cdot{\bf{a_1}}} + e^{i\bf{k}\cdot{\bf{a_2}}}) + t_3 
   (e^{ i{\bf k}\cdot( {\bf a}_1 - {\bf a}_2) } \nonumber \\
 &+& e^{ i{\bf k}\cdot( {\bf a}_2 - {\bf a}_1) }
 + e^{ i{\bf k}\cdot( {\bf a}_1 + {\bf a}_2) })
 \end{eqnarray}
\end{itemize}
The two other matrix elements i.e.
$\langle 0 |c_{k\sigma^\prime} \rm{H} c^\dagger_{k\sigma}|0\rangle$, $\langle 0 |d_{k\sigma^\prime} \rm{H} d^\dagger_{k\sigma}|0\rangle$ remain the same as the second-nearest neighbor case. 

Now we are in a position to construct all the matrix elements of h-BN and monolayer graphene using interactions up to the third nearest neighbor.

\underline{\textbf{h-BN}}
\begin{eqnarray}
\sum_p \frac{\epsilon^A_p}{N} \delta_{\sigma, \sigma^\prime} &=& t_b  \nonumber \\
\sum_p \frac{\epsilon^B_p}{N} \delta_{\sigma, \sigma^\prime} &=& t_N \nonumber \\
\rm{H_1} &=& t_1\left(1 + e^{i{\bf k}\cdot{\bf a}_1} + e^{i{\bf k}\cdot{\bf a}_2} \right) \nonumber \\
\rm{H_2} &=& t_2\left( e^{i{\bf k}\cdot{\bf a}_1} + e^{i{\bf k}\cdot{\bf a}_2} + e^{i{\bf k}\cdot({\bf a}_1 - {\bf a}_2)}
+ e^{i{\bf k}\cdot({\bf a}_2 - {\bf a}_1)} + e^{-i{\bf k}\cdot{\bf a}_1} + e^{-i{\bf k}\cdot{\bf a}_2}
\right) \nonumber \\             
\tilde{\rm{H_2}} &=& \tilde{t_2}\left( e^{i{\bf k}\cdot{\bf a}_1} + e^{i{\bf k}\cdot{\bf a}_2} + e^{i{\bf k}\cdot({\bf a}_1 - {\bf a}_2)}+ e^{i{\bf k}\cdot({\bf a}_2 - {\bf a}_1)} + e^{-i{\bf k}\cdot{\bf a}_1} + e^{-i{\bf k}\cdot{\bf a}_2}
\right) \nonumber \\
\rm{H_3} & =& t_3 \left(
e^{ i{\bf k}\cdot( {\bf a}_1 - {\bf a}_2) } 
+ e^{ i{\bf k}\cdot( {\bf a}_2 - {\bf a}_1) }
+ e^{ i{\bf k}\cdot( {\bf a}_1 + {\bf a}_2) } 
\right)  \nonumber
\end{eqnarray}

\begin{gather*}
    \rm{H} = \begin{bmatrix}
    t_b + \rm{H}_2 & 0 & 0 & \rm{H}_1 + \rm{H}_3\\
    0 & t_b + \rm{H}_2 & \rm{H}_1 + \rm{H}_3 & 0\\
    0 & \rm{H}_1^\dagger + \rm{H}_3^\dagger & t_n + \tilde{\rm{H}_2} & 0\\
    \rm{H}_1^\dagger + \rm{H}_3^\dagger & 0 & 0 & t_n + \tilde{\rm{H}_2}
    \end{bmatrix} \,,
\end{gather*}
\begin{table}[h!]

\caption{Tight binding parameters for h-BN}
\begin{center}
    \begin{tabular}{||p{2cm}|p{2cm}|p{2cm}|p{2cm}|p{2cm}||}
     \hline
     $t_b$ (eV)& $t_n$ (eV)& $t_1$ (eV)& $t_2$ (eV)& $t_3$ (eV)\\
     \hline
     \hline
     2.46   & -2.55    & 2.16 &   0.04 & 0.08\\
     \hline
    \end{tabular}
\end{center}
\label{Table1}
\end{table}

\underline{\textbf{Monolayer graphene}}
\begin{eqnarray}
\sum_p \frac{\epsilon^A_p}{N} \delta_{\sigma, \sigma^\prime} &=& t_C  \nonumber \\
\rm{H_1} &=& t_1\left(1 + e^{i{\bf k}\cdot{\bf a}_1} + e^{i{\bf k}\cdot{\bf a}_2} \right) \nonumber \\
\rm{H_2} &=& t_2\left( e^{i{\bf k}\cdot{\bf a}_1} + e^{i{\bf k}\cdot{\bf a}_2} + e^{i{\bf k}\cdot({\bf a}_1 - {\bf a}_2)}
+ e^{i{\bf k}\cdot({\bf a}_2 - {\bf a}_1)} + e^{-i{\bf k}\cdot{\bf a}_1} + e^{-i{\bf k}\cdot{\bf a}_2}
\right) \nonumber \\             
\rm{H_3} & =& t_3 \left(
e^{ i{\bf k}\cdot( {\bf a}_1 - {\bf a}_2) } 
+ e^{ i{\bf k}\cdot( {\bf a}_2 - {\bf a}_1) }
+ e^{ i{\bf k}\cdot( {\bf a}_1 + {\bf a}_2) } 
\right)  \nonumber
\end{eqnarray}

\begin{gather*}
    \rm{H} = \begin{bmatrix}
    t_C + \rm{H}_2 & 0 & 0 & \rm{H}_1 + \rm{H}_3\\
    0 & t_C + \rm{H}_2 & \rm{H}_1 + \rm{H}_3 & 0\\
    0 & \rm{H}_1^\dagger + \rm{H}_3^\dagger & t_C + \rm{H}_2 & 0\\
    \rm{H}_1^\dagger + \rm{H}_3^\dagger & 0 & 0 & t_C + \rm{H}_2
    \end{bmatrix} \,,
\end{gather*}

In the case of graphene, we also model electronic interaction between opposite spins through a Hubbard Hamiltonian with the repulsion parameter being denoted by $U$. Since the average number of electrons with spin-up is taken as 1 and the average number of electrons with spin-down is taken as 0. Therefore, $U$ enters the Hamiltonian only on the down-spin diagonal terms.

\begin{align}
    \rm{H} = \begin{bmatrix}
    t_C + \rm{H}_2 & 0 & 0 & \rm{H}_1 + H_3\\
    0 & t_C + \rm{H}_2 + $\textit{U}$ & \rm{H}_1 + \rm{H}_3 & 0\\
    0 & \rm{H}_1^\dagger + H_3^\dagger & t_C + \rm{H}_2 + $\rm{\textit{U}}$ & 0\\
    \rm{H}_1^\dagger + \rm{H}_3^\dagger  & 0 & 0 & t_C + \rm{H}_2
    \end{bmatrix} 
\end{align}

\begin{table}[h!]
\caption{Hubbard model parameters for graphene}
\begin{center}
    \begin{tabular}{||p{2cm}|p{2cm}|p{2cm}|p{2cm}|p{2cm}||}
     \hline
     $t_C$ (eV)& $t_1$ (eV)& $t_2$ (eV)& $t_3$ (eV) & $U$ (eV)\\
     \hline
     \hline
     1.994    & 2.86 &   -0.236 & 0.252 & 9.3\\
     \hline
    \end{tabular}
\end{center}
\label{Table2}
\end{table}

\section{Scaling}
After the single qubit rotations ($R_y$) and before the Controlled-Controlled Rotations ($C-C-R$), the probability distribution corresponding to a specific ${\sigma^z, h}$ can be written as:
\begin{equation}
    \frac{e^{\frac{1}{k}(\sum_{i}a_i\sigma^z_i + \sum_{j}b_j h_j)}}{\sum_{{\sigma^z, h}}e^{\frac{1}{k}(\sum_{i}a_i\sigma^{z}_i + \sum_{j}b_j h_j)}}\\[11pt]
\end{equation}

Once the $C-C-R$ is applied for the first time, the probability distribution with the corresponding ancilla qubit being in $\ket{1}$ is:

\begin{equation}
    \frac{e^{\frac{1}{k}(\sum_{i}a_i\sigma^z_i + \sum_{j}b_j h_j)}}{\sum_{{\sigma^z, h}}e^{\frac{1}{k}(\sum_{i}a_i\sigma^{z}_i + \sum_{j}b_j h_j)}} \times \frac{e^{\frac{1}{k}(w_{ij}\sigma^z_ih_j)}}{e^{\frac{1}{k}|w_{ij}|}}
\end{equation}

After all the $C-C-R$ are applied, then the probability distribution with all the ancilla qubits being in $\ket{1}$ is:
\begin{equation}
     \frac{e^{\frac{1}{k}(\sum_{i}a_i\sigma^z_i + \sum_{j}b_j h_j)}}{\sum_{{\sigma^z, h}}e^{\frac{1}{k}(\sum_{i}a_i\sigma^{z}_i + \sum_{j}b_j h_j)}} \times \prod_{i,j} \frac{e^{\frac{1}{k}(w_{ij}\sigma^z_ih_j)}}{e^{\frac{1}{k}|w_{ij}|}} \nonumber
\end{equation}
\begin{equation}
     =\frac{e^{\frac{1}{k}(\sum_{i}a_i\sigma^z_i + \sum_{j}b_j h_j)}}{\sum_{{\sigma^z, h}}e^{\frac{1}{k}(\sum_{i}a_i\sigma^{z}_i + \sum_{j}b_j h_j)}} \times  \frac{e^{\frac{1}{k}(\sum_{i,j}w_{ij}\sigma^z_ih_j)}}{e^{\frac{1}{k}\sum_{i,j}|w_{ij}|}} \nonumber
\end{equation}
\begin{equation}
     =\frac{e^{\frac{1}{k}(\sum_{i}a_i\sigma^z_i + \sum_{j}b_j h_j + \sum_{i,j}w_{ij}\sigma^z_ih_j)}}{\sum_{{\sigma^z, h}}e^{\frac{1}{k}(\sum_{i}a_i\sigma^{z}_i + \sum_{j}b_j h_j)}e^{\frac{1}{k}\sum_{i,j}|w_{ij}|}} \label{scaling_equation_1}
\end{equation}

By summing all possible states, the probability of getting all ancilla qubits to be in $\ket{1}$ is given by:
\begin{equation}
     \frac{\sum_{\sigma^z,h}e^{\frac{1}{k}(\sum_{i}a_i\sigma^z_i + \sum_{j}b_j h_j + \sum_{i,j}w_{ij}\sigma^z_ih_j)}}{\sum_{{\sigma^z, h}}e^{\frac{1}{k}(\sum_{i}a_i\sigma^{z}_i + \sum_{j}b_j h_j)}e^{\frac{1}{k}\sum_{i,j}|w_{ij}|}} \label{scaling_equation_2}
\end{equation}

Since, $e^{w_{ij}\sigma^z_ih_j} \geq e^{-|w_{ij}|}$, the term $e^{w_{ij}\sigma^z_ih_j}$ in Eq.(\ref{scaling_equation_2}) can be replaced with $e^{-|w_{ij}|}$.

This results in the successful probability P to be:
\begin{eqnarray}
    P &=& \frac{\sum_{\sigma^z,h}e^{\frac{1}{k}(\sum_{i}a_i\sigma^z_i + \sum_{j}b_j h_j + \sum_{i,j}w_{ij}\sigma^z_ih_j)}}{\sum_{{\sigma^z, h}}e^{\frac{1}{k}(\sum_{i}a_i\sigma^{z}_i + \sum_{j}b_j h_j)}e^{\frac{1}{k}\sum_{i,j}|w_{ij}|}} \geq \frac{e^{\frac{-1}{k}(\sum_{ij}|w_{ij}|)}}{e^{\frac{1}{k}(\sum_{ij}|w_{ij}|)}} =  \frac{1}{e^{\frac{1}{k}(\sum_{ij}2|w_{ij}|)}}
\end{eqnarray}

By choosing $\max(\sum_{ij}\frac{|w_{ij}|}{2}, 1)$, the lower bound of the probability for successful sampling becomes a constant equal to $e^{-4}$.
\section{Implementation Details}
\begin{enumerate}
\item If n is the number of visible units and m is the number of hidden units, then,
\begin{enumerate}
\item The number of qubits required are:
\begin{itemize}
    \item 2 qubits for visible units (n)
    \item 2 qubits for hidden units (m)
    \item 4 ancilla qubits (n+m)
\end{itemize}
\item The number of gates used are:
\begin{itemize}
    \item 4 single qubit rotations (n+m)
    \item 4 Controlled-Controlled rotations (n$\times$m)
    \item 24 X(bit-flip) gates (6$\times$n$\times$m)
\end{itemize}
\end{enumerate}
\item The parameter are updated through gradient descent with a learning rate equal to 0.01.
\item The number of measurements 
\[   = \text{number of iterations = }
\begin{cases}
    \approx 30000,& \text{for classical and $qasm$ simulations without warm start}
    \\
    \approx 500,& \text{for classical and $qasm$ simulations with warm start}
    \\
    \approx 500, & \text{for IBM-Q implementation}
\end{cases}
\]

\end{enumerate}

\section{Result (without a warm start or measurement error mitigation)}

\begin{figure}[H]
    \centering
    \includegraphics[width=0.75\textwidth]{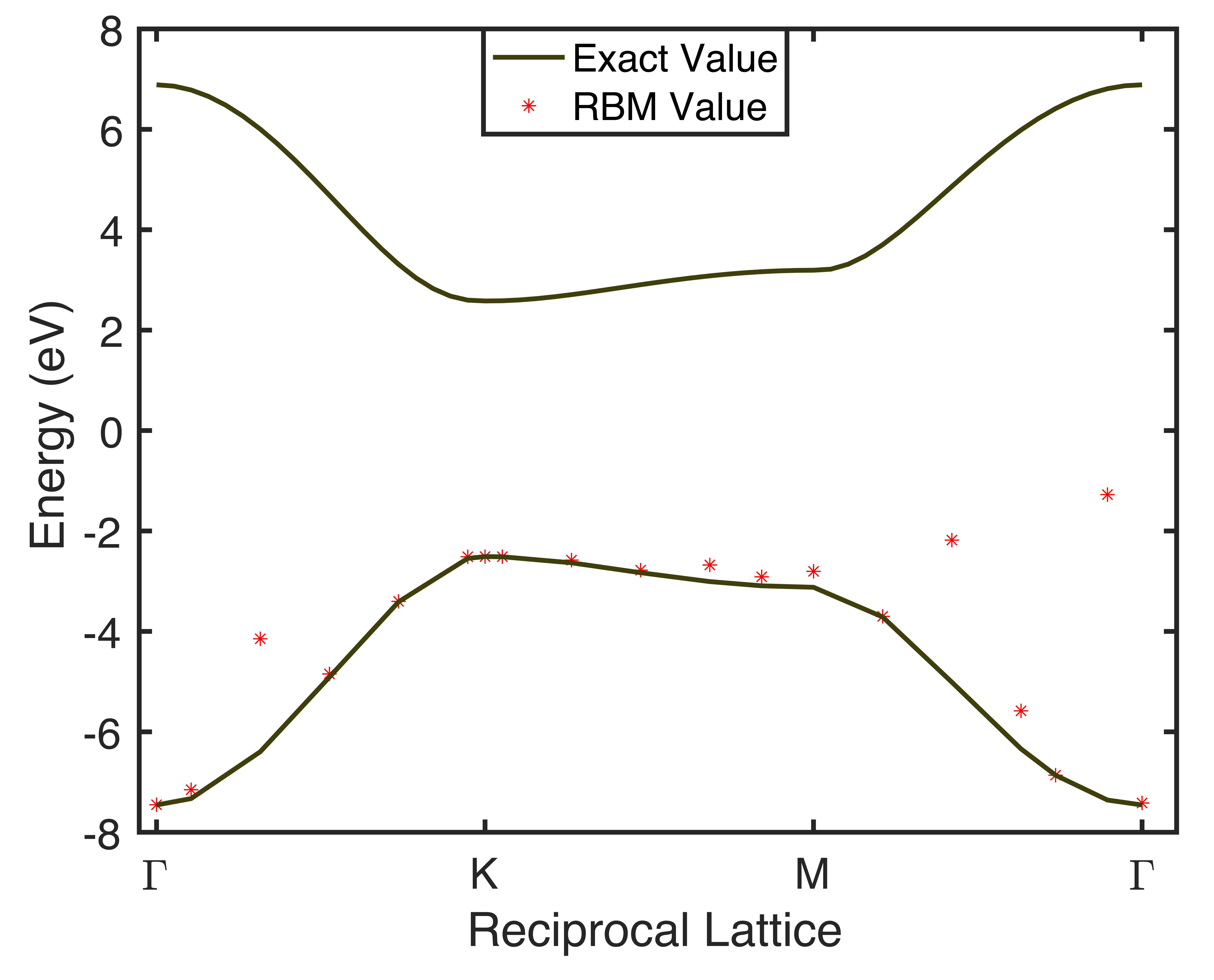}
    \caption{ Band structure of h-BN without a warm start or measurement error mitigation}
    \label{hbn_actual_without_TF}
\end{figure}

The result corresponding to h-BN without providing a warm start and without employing measurement error mitigation is shown in Fig. \ref{hbn_actual_without_TF}

\end{document}